\newcommand{\gcas}{\object{$\gamma$\,Cas}}
\newcommand{\hd}{\object{HD\,110432}}

\documentclass{aa} \usepackage{natbib,graphicx}  \usepackage{natbib,graphicx} 

\begin{document}


\title{$\gamma$ Cassiopeiae: an X-ray Be star with personality\thanks{This work is based on observations obtained with XMM-{\it 
Newton}, an ESA science mission with instruments and contributions directly funded by ESA Member
States and NASA.
}}

\author{R. Lopes de Oliveira\inst{1} \and M. A. Smith\inst{2} \and C. Motch\inst{3}}

\offprints{R. Lopes de Oliveira,\\
\email{rlopes@astro.iag.usp.br}}

\institute{ 
Instituto de Astronomia, Geof\'{\i}sica e Ci\^encias Atmosf\'ericas, 
Universidade de S\~ao Paulo, R. do Mat\~ao 1226, 05508-090 S\~ao Paulo, Brazil 
\and Catholic University of America, 3700 San Martin Drive, Baltimore, MD 21218, USA
\and Observatoire Astronomique, UMR 7550 CNRS, 11 rue de l'Universit\'e, F-67000 Strasbourg, France
}
\date{Received 10 November 2008 / Accepted 6 January 2010}

\authorrunning{R. Lopes de Oliveira}  \titlerunning{\gcas: an X-ray Be star with personality}

\abstract{

An exciting unsolved problem in the study of high energy processes of
early type stars concerns the physical mechanism for
producing X-rays near the Be star $\gamma$ Cassiopeiae. 
By now we know that this source and several ``\gcas\   analogs" exhibit 
an unusual hard thermal X-ray spectrum, compared both to normal
massive stars and the non-thermal emission of known Be/X-ray binaries. 
Also, its light curve is variable on almost all conceivable timescales.
In this study we reanalyze a high dispersion spectrum obtained by
{\it Chandra} in 2001 and combine it with the analysis of a new (2004) spectrum 
and light curve obtained by XMM-{\it Newton.} We find that both spectra
can be fit well with 3--4 optically thin, thermal 
components consisting of a hot component having a temperature 
$k$T$_Q$ $\sim$ 12--14\,keV, {\it perhaps} one with a value of $\sim$ 2.4\,keV, 
and two with well defined values near 0.6\,keV and 0.11\,keV.
We argue that these components arise in discrete (almost monothermal) 
plasmas. Moreover, they cannot be produced within an integral gas structure
or by the cooling of a dominant hot process. 
Consistent with earlier findings, we also find that the Fe
abundance arising from K-shell ions is significantly subsolar 
and less than the Fe abundance from L-shell ions.
We also find novel properties not present in the
earlier {\it Chandra} spectrum, including a dramatic decrease in the 
local photoelectric absorption of soft X-rays, a decrease in the
strength of the Fe and possibly of the Si K fluorescence features, underpredicted lines in
two ions each of Ne and N (suggesting abundances that are $\sim$ 1.5--3$\times$ 
and $\sim$ 4$\times$ solar, respectively),
and broadening of the strong Ne\,X\,Ly$\alpha$ and 
O\,VIII\,Ly$\alpha$ lines.
In addition, we note certain traits in the \gcas\ spectrum that 
are different from those of the fairly well studied analog \hd\ 
- in this sense the stars have different ``personalities."
In particular, for \gcas\ the hot X-ray component remains
nearly constant in temperature, and the photoelectric absorption of the
X-ray plasmas can change dramatically. As found by previous investigators of
\gcas, changes in flux, whether occurring slowly or in rapidly evolving 
flares, are only seldomly accompanied by variations in hardness. Moreover,
the light curve can show a ``periodicity" that is due to the presence of
flux minima that recur semiregularly over a few hours, and which can 
appear again at different epochs. 

}

\maketitle

\section{Introduction}\label{introduction}

The observed properties of \gcas\ (B0.5\,Ve; m$_{V}$\,=\,2.25) 
in ultraviolet to infrared wavelengths have
led to major discoveries related to 
the Be-phenomenon since its discovery as the 
first of what became known as ``Be stars" \citep{Secchi67}. 
However, its X-ray emission is not typical of this kind of object \citep{W82}.
The X-ray emission of \gcas\ is dominated by a hot plasma component 
($k$T $\sim$ 10--12\,keV), with mildly high luminosity 
($\sim$\,10$^{32-33}$\,erg\,s$^{-1}$) and variable flux \citep[][and 
references therein]{Smith04}.
Be stars at large are softer X-ray emitters ($k$T $\sim$ 0.5\,keV) with 
lower luminosity ($\la$\,10$^{32}$\,erg\,s$^{-1}$) and little variability 
\citep[e.g.,][]{Berghofer97}. 
In contrast, most of the well investigated
Be/X-ray binary systems are Be+neutron star systems and show a high 
luminosity ($\ga$ 10$^{33}$\,erg\,s$^{-1}$) and a distinctly nonthermal high
energy distribution. 
Notably, no large X-ray outburst has been
observed in \gcas. This is at variance 
with the behavior witnessed in many classical Be/X-ray systems.
\gcas\ is now known to be in a 204-day binary, with an
eccentricity variously determined to be near $e$ = 0 and 0.26 
\citep{Harmanec00,Miros02}.
Little is known about the secondary, except
that it is likely to have a mass in the range 0.5--2 M$_{\rm \odot}$.
Recently, X-ray properties similar to those of \gcas\ have been observed 
in a small but growing number of Be stars: 
the \gcas-like stars \citep{Motch07,Lopes06,Lopes07T}. 

A conclusive explanation for the X-ray emission of \gcas\ and 
its analogs is lacking.
The first suggestion to explain the hard X-ray emission of \gcas\ was that it is powered by accretion of matter from the Be wind or disk onto
an accreting neutron star \citep{W82}.  More recently, some have advocated the
presence of an accreting white dwarf companion. This suggestion
comes from several X-ray characteristics of cataclysmic variable systems 
\citep[e.g.,][]{M86,K98},
such as their thermal nature and only moderate X-ray luminosity. However,
the analogy is incomplete upon further scrutiny, and X-rays would have 
to come from accretion onto a white dwarf representative of 
a new class of Be/X-ray binaries \citep{Lopes07} and have a high
efficiency of mass accretion energy to X-ray flux.

A second proposed scenario suggests that the X-ray emission of \gcas\ is 
a consequence of magnetic interaction between its stellar surface and a 
Be (Keplerian decretion) circumstellar disk that entrains a magnetic field. This idea 
is based in part on the correlations of the UV and optical variabilities 
with the X-ray light curve and also indirect evidence of magnetic field 
observed in this star. This evidence takes the form
of migrating subfeatures running blue-to-red through optical 
and UV line profiles \citep{Yang1988, SRH98} 
and also the discovery of
a gray, robust, 1.21581-day feature in the star's light curve. 
This periodicity must be very near or identical to the 
star's rotation period \citep{SHV06}. 
The inference of a disk connection is
based on the cyclical aspect and reddish tinge of the optical cycles that
correlate with the X-ray ones. It is also supported by patterns of
variability in the C\,IV and Si\,IV lines that are occasionally seen in the
X-ray light curve \citep{C00}.
One piece of evidence for this interaction comes from the 
observation of highly redshifted {\it absorption} UV lines,
suggestive of material being ejected from the circumstellar environment 
toward the star with enough energies to produce $\sim$ 10\,keV X-rays when 
they impact the star \citep{SR99}.

Both the magnetic disk and accretion 
interpretations have important astrophysical implications.
The former suggests that disk dynamos 
operate and also that these stars are possible proto-magnetars. The
accretion interpretation would imply the presence of a neutron star
in an unusual accretion regime, or a white dwarf  with novel
properties. The presence of
a white dwarf is still speculative, but such objects are predicted 
to be common as secondaries in models 
of evolution of binaries with B primaries.  
The resolution of the mystery of the production of X-rays in 
``\gcas\ stars" could lead to a breakthrough in either of these fields.

We have focused our efforts to try to understand the origin of the 
X-rays of \gcas\ and analog Be stars. In this paper we 
confine ourselves to the X-ray 
properties of \gcas\ from high and medium resolution spectroscopy and 
timing studies using data obtained by XMM-\textit{Newton} satellite. 
{\it Chandra} HETG data are also reinvestigated.

\section{Previous X-ray observations of $\gamma$\,Cas}

 Several {\it Rossi X-ray Timing Explorer (RXTE)} observations of the light 
curve of \gcas\ detailed by
\citet{SRC98} and \citet{RS00} (hereafter ``SRC98" and ``RS00," 
respectively) have disclosed that the light curve undergoes
variations on rapid (flaring), intermediate (several hours) and long (2-3 
month cycles) timescales.  We summarize the flaring results first as follows:

\begin{itemize}

\item Flares (shots) are ubiquitous, except during brief
 periods of very low X-ray flux. The reoccurrence of these lulls
is often cyclical. 

\item Individual flare profiles are narrow and symmetrical in shape.

\item Collectively flares show a log-normal distribution in energy.

\item The flares show an approximately $1/f$ distribution down to the
 photon limit of the instrument (for {\it RXTE} about 4 seconds). 
 Occasional groups of flares (aggregates) last as long as a few minutes,
and indicate the end of this red noise dependence.

\item Spectrophotometry of shots indicates that their flux is usually
 difficult to distinguish from that of the underlying basal flux.
 That is, the two components have about the same high temperature of 
 10--12\,keV.  

\end{itemize}

 Apparently random variations of several hours and 
10's of percent amplitude occur often in the \gcas\ X-ray light curve.
One long simultaneous observation with the {\it Goddard High Resolution
Spectrograph} (GHRS) on the {\it Hubble Space Telescope} established the
correlation of small-amplitude UV continuum fluctuations \citep{SRC98,SR99} 
and the concomitant absorptions in either lines arising from either 
lower or higher than expected ionization and excitation states (SR99).

Long-term variations have also been reported with 
cycle lengths of 50-91 days and amplitudes of a factor of three
(Robinson, Smith, \& Henry 2002, ``RSH02").
These are so far well correlated with  corresponding cycles in optical 
bands (Johnson $B$ and $V$). Although the optical amplitudes are 100 times 
smaller than the X-ray ones they are still much larger than in terms of
luminosity output and therefore cannot be the result of reprocessing of 
X-rays.  Rather, a common mechanism appears to mediate both variabilities.

Modulations with several hours timescales were suggested by previous X-ray satellites.
\citet{Frontera87} reported a modulation with timescale of $\sim$ 1.67 hr from EXOSAT observation
on 1984 December 7, while \citet{P93} argued from a reanalysis of this data that such 
an oscillation arise from statistical fluctuations in the red noise spectrum of the source. 
Also, \citet{P93} have not found periodic oscillation in other EXOSAT data obtained on 1984 December 25-26.
A $\sim$ 2.3\,hr oscillation was suspected by \citet{Haberl95} from 
ROSAT observation on 1993 July 16-17.
\citet{O99} found weak evidence for a $\sim$ 0.6\,h and 2.6\,hr in 
BeppoSAX observation of \gcas\ carried out on 1998 July 20-21. 
None of these suspected detections could be found by subsequent monitorings
of the star with the RXTE (e.g., Smith, Robinson, \& Henry 2000).

 \begin{figure*} \centering{ 
\includegraphics[bb=1.8cm 12.5cm 19.9cm 22.5cm,clip=true,width=18cm]{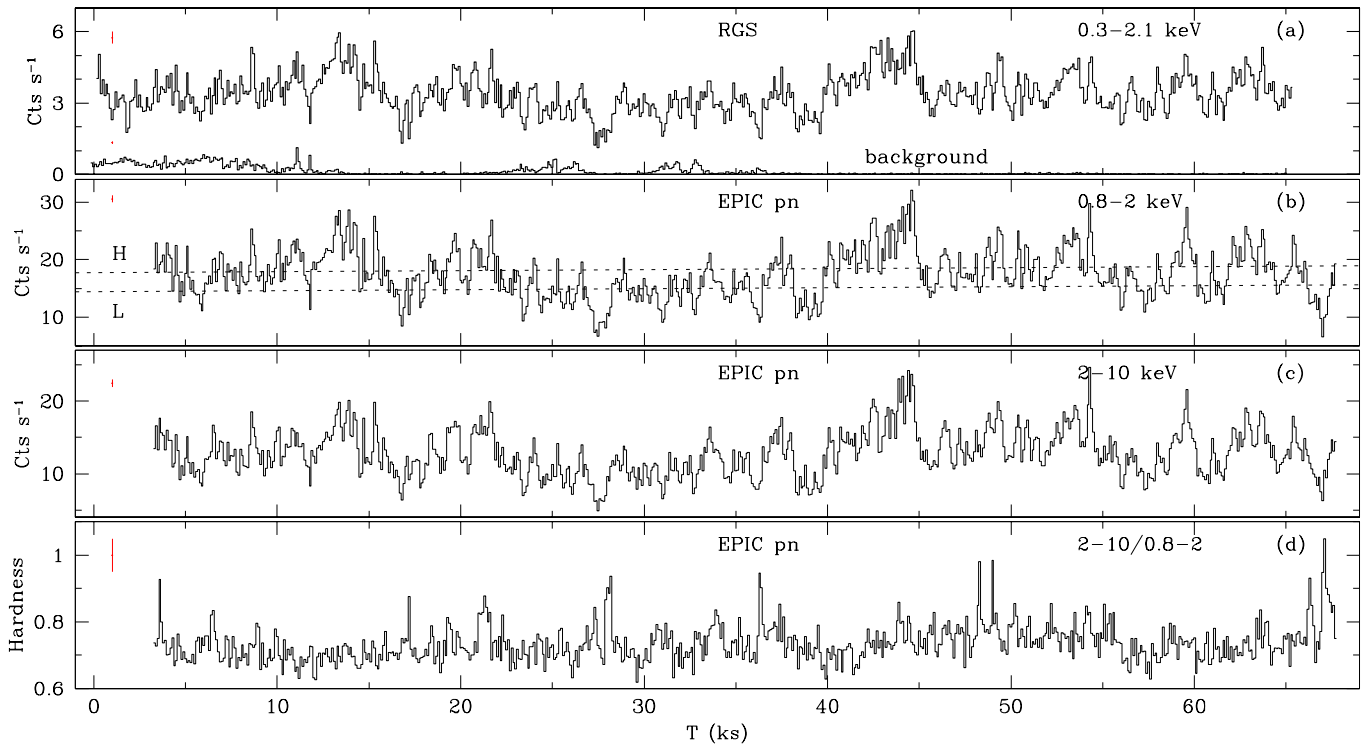}
\caption{RGS1+RGS2 light curve combining 1st and 2nd orders and the RGS1 background light curve (a), and EPIC $pn$ light curves in the 0.8--2\,keV (b) and 2--10\,keV (c) X-ray bands, and the respective hardness ratio (d). Time bins of 100 s. 
Upper limits for error bars at 1\,$\sigma$ are plotted in the left corner of the figures. The horizontal lines in panel (b) represent the thresholds adopted to define the high (H) and low (L) state spectra (see Section \ref{sct:fekcomplex}).
\label{fig:lc_hrd_gcas}}} 
\end{figure*}

Pre-high resolution spectroscopy of \gcas\ has demonstrated that its X-ray
emission is dominated by a thermal component  with $k$T $\sim$ 10--12\,keV 
\citep{M86,P93,K98,
O99}. These spectra exhibited Fe\,XXV and Fe\,XXVI\,Ly$\alpha$ K line strengths consistent 
with this temperature, but they are consistent only with a 
subsolar Fe abundance of 0.2--0.4\,Z$_{\odot}$.

Prior to this paper, the only high dispersion spectrum of \gcas\ was obtained
on 2001 August 10
by the {\it Chandra} High Energy Transmission Grating (HTEG) discussed
by \citet{Smith04} (hereafter ``S04"). This spectrum
showed a complex structure caused by plasma radiating in at least 3--4
temperatures ranging from about 12\,keV to about 0.15\,keV. 
The hot component in turn consisted
of two subcomponents with very different ($\sim$10$^{23}$ cm$^{-2}$ and 
3$\times$10$^{21}$ cm$^{-2}$) column densities. While the Fe\,K
lines again indicated abundances of only 0.22${\pm 0.05}$\,Z$_{\odot}$ 
other Fe lines arising from the L-shell give abundances which were consistent 
with the solar value. The strengths of other lines were also consistent 
with solar abundances.

\section{Observations and data reduction}

\gcas\ was observed by the XMM-\textit{Newton} X-ray observatory on 2004 February 5 during about 
68\,ks in the revolution 762 (ObsId 0201220101). 
This observation was performed 
with the EPIC $pn$ camera running in the {\it fast timing mode} (timing resolution
of 0.03 ms), connected to the 
{\it thick} optical filter,  and with
the high spectral resolution RGS1 and RGS2 cameras. The optical monitor 
was blocked as usual for bright optical sources, and the central CCDs in the 
MOS1/2 cameras were switched off in order to avoid overloading the telemetry.
Data reduction has been made using the Science Analysis System (SAS)
software v8.0.1. All data were reprocessed using the pipeline \textsc{epproc} 
(for EPIC $pn$ camera) and \textsc{rgsproc} (for RGS cameras)
tasks. For the timing analysis, we use the Z$^{2}_{n}$ Rayleigh \citep{Buccheri83} 
and the Scargle/Midas \citep{Scargle82} peridograms, and the 
Xronos\footnote{http://heasarc.nasa.gov/docs/xanadu/xronos/xronos.html} package.
For spectral fits, the 
XSPEC\footnote{http://heasarc.nasa.gov/docs/xanadu/xspec/index.html} software 
v11.3.2 was applied.
The EPIC $pn$ data obtained in the fast timing mode have a doubtful calibration 
and increased noise especially at softer energies \citep[$<$ 0.5\,keV;][]{Guainazzi08} 
and, to be on the safe side, timing analysis was restricted to broadbands at the 
0.8--10\,keV energy range without relevant loss of informations. 
For spectral studies, since we have high quality spectra from 
RGS1 and RGS2 cameras covering the soft X-rays (6.2--38\AA; 
or $\sim$ 0.3--2\,keV), we use the EPIC $pn$ data as complementary data 
covering the hard part of the spectrum (1.25--4.1\AA; or $\sim$ 3--10\,keV).
We noticed that the inclusion of the low energy part of the EPIC $pn$ 
data produces systematic residuals that can be connected to calibration uncertainties.
Because of the high count rate of \gcas, the inclusion of times with slight 
background flares during the XMM-{\it Newton} observation has little impact
on its EPIC $pn$ and RGS light curves. However, we opted for excluding 
these times in the spectral analysis. 
The resulting exposure times in the flare-free good time intervals were 48\,ks 
for the RGS1 and RGS2 cameras and 51.2 ks for EPIC $pn$.

We reinvestigated the {\it Chandra} data of \gcas\ obtained on 2001 
August 10 with the HETG during about 52\,ks (ObsID 1895) with the CIAOv4.0, 
reported by \citet{Smith04}. Our purpose 
was to compare the {\it Chandra} and XMM-{\it Newton} spectra 
of \gcas\ from the same analysis techniques and thus to insure that
any differences between the derived parameters are not due to the model
fitting programs. 
Our reanalysis of the {\it Chandra} spectrum with XSPEC and families of 
{\it mekal} models
gives results consistent with those reported by S04, obtained from Sherpa 
and APEC models.

\section{Timing studies}

The soft (0.8--2\,keV) and hard (2--10\,keV) light curves of \gcas\ are 
marked for strong variabilities on timescales ranging from a few to 
thousand of seconds 
(Fig. \ref{fig:lc_hrd_gcas}). 
On long timescales the variation attains $\sim$80\% of the mean flux.
Superposed on this global modulation are ubiquitous flare-like events.
We note in particular that the soft X-ray light curve responds faithfully
to the rapid flares of the high energy light curve.
In addition, the hardness of the source (Fig. \ref{fig:lc_hrd_gcas}c)
usually exhibits nearly the same variations as the integrated flux.
For example, the formal slope of the hardness-total count rate curve
from our observations (not shown) is only (5.2$\pm$2.9)$\times$10$^{-4}$.

\begin{figure} \centering{ 
\includegraphics[bb=1.3cm 15.2cm 12.4cm 24.5cm,clip=true,width=9cm]{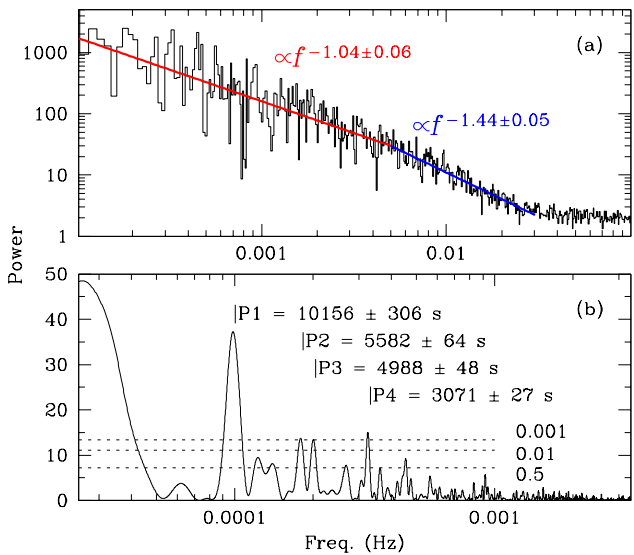}}
\caption{ 
Power spectrum. Top: from EPIC $pn$ events at 0.8--10\,keV, rebinned as a 
geometrical series.  Bottom: from 0.8--10\,keV EPIC $pn$ data binned to 100 s; 
the dashed lines represent the confidence levels.
\label{fig:spctpotgcas}} 
\end{figure}
 
These findings are similar to those reported 
by SRC98, though RS00 found that shot fluxes tended to 
exhibit a lower (softer) ratio.

Figure \ref{fig:spctpotgcas}a exhibits the power spectrum 
of the EPIC $pn$ events from the combined spectral bands
shown in Fig.\,\ref{fig:lc_hrd_gcas}. A key point from this
figure is that at frequencies above about 0.003\,Hz
the slope of the distribution is significantly steeper than -1. 
This was also noted by SRC98 and RS00, who found slopes of
-1.23 and -1.36 in 1996 and 1998, respectively.
The difference between those values was already marginally statistically
significant, and the current result is certainly even more significantly
different from the 1996 result. As noted by RS00, the difference
between these slopes and -1 is most likely due to a relative prevalence of 
strong, longer-lived flares as compared to rapid short ones in these data. 
RS00 reported $1/f$ slopes in their 1998 light curves that were 
intermediate between the 1996 slope of -1 and the present steeper one.
The explanation for the knee at about 0.003\,Hz is again likely 
to be caused by the absence of discrete flares with timescales longer
than a few minutes. 

\begin{figure}[t!]\centerline{
\includegraphics[bb=1cm 19.2cm 12.5cm 24.48cm,clip=true,width=8.8cm]{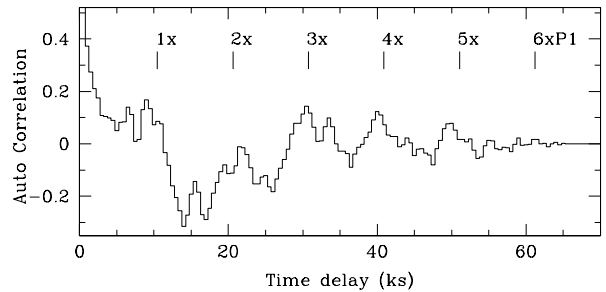}}
\caption{Autocorrelation from 0.8--10\,keV light curve binned to 500 s. 
P1 refers to result of Fig. \ref{fig:spctpotgcas}b.
\label{fig:gcas_autocor}}  
\end{figure} 

The break is due to the dominance of apparently
random variations that often occur in the X-ray light curve on 
timescales of about a half hour or longer.
At least some of these appear to be connected with variations in the
UV continuum, which \citet{SRC98} have associated with the partial 
occultation of the star by rotationally advected translucent corotating
clouds.

\begin{figure*}[t!]
\centering
\includegraphics[bb=1.9cm 2.5cm 14.5cm 27.2cm,clip=true,angle=-90,width=18cm]{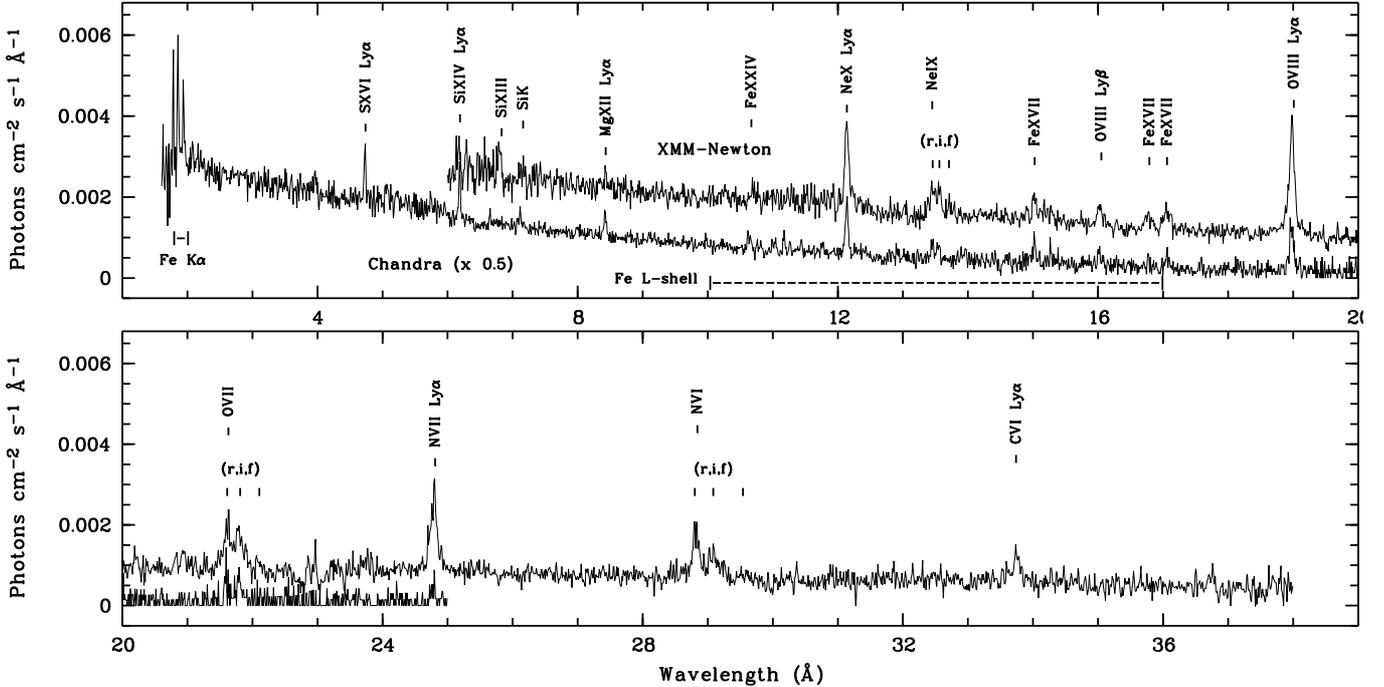}
\caption{The high resolution fluxed spectrum of \gcas, from the RGS1+RGS2 XMM-{\it Newton} cameras combining 1st and 2nd orders, and from the HEG+MEG Chandra cameras combining all four first-order spectra. \label{fig:spctgcas_rgs12_flux}}
\end{figure*}

Another interesting feature emerging from the autocorrelation 
analysis of a light curve obtained by the RXTE satellite
 (Fig.\,\ref{fig:gcas_autocor}) 
was the reappearance of periodic flux lulls.
Following the procedure of \citet{RS00}
we have cross-correlated both the fluxes in our XMM-{\it Newton}
light curve and also the reciprocals of these fluxes. The latter exercise was 
carried out because RS00 had found that
the autocorrelation of the direct fluxes produced no distinguishable
features, whereas the autocorrelation of the reciprocal fluxes produced 
dramatic variations with lags of spacing of 7.5 
hours and higher multiples. These authors reasoned that this different 
behavior could come about if X-ray active centers at various latitudes of 
the rotating star would not produce marked features in the autocorrelation
curve. However, features would be observed from the curve generated from
the reciprocal fluxes if only a few longitudes of the star were not covered
with X-ray activity. Our results, shown in Fig.\,\ref{fig:gcas_autocor} 
for the direct flux, produced rather similar curves for both
the direct and reciprocal fluxes, indicating that the distribution of
X-ray centers had rearranged themselves dramatically between the two epochs
of observation. The curve itself shows evenly spaced maxima every 
$\sim$ 10\,156\,s (2.8 hours; Fig. \ref{fig:spctpotgcas}b). 
On a subsequent study of six {\it RXTE} light curves of \gcas, \citet{RSH02}
discovered that autocorrelation peaks were most apparent in light
curves associated with flare fluxes. These authors found that the
features are centered at different time lags at different epochs, and
occasionally did not occur at all.  At various times these 
authors found spacings at 7.5 hours, 3.5 hours, and 5.8 hours. 
The 2.8 hours spacing we now find appear to be a recurrence of the same 
cycle length these authors found during a long RXTE 
observation in 2000 December.  
As \citet{RSH02} found, the autocorrelation peaks seem to
be caused by an absence of flux (lulls) at different times. As such,
they seem to hint at the existence of X-ray relaxation cycles in the \gcas\
environment. In contrast, \citet{Lopes07} found no such lulls in the
light curve of \hd.

\section{Spectroscopic properties}
 
\subsection{Comparison with previous {\it Chandra} HETGS spectrum}

 We have obtained a XMM-{\it Newton} spectrum in order 
to investigate time-dependent 
differences and also to exploit the higher effective aperture of the system, 
albeit with lower resolution at higher energies. 
The XMM-{\it Newton} provides coverage over longer wavelengths 
($\sim$ 1--38\AA) than an {\it Chandra} spectrum 
with the high and medium energy gratings ($\sim$ 1.5--25\AA).

 Our XMM-{\it Newton} spectrum shows the same general properties as the 
2001 {\it Chandra} spectrum (Fig. \ref{fig:spctgcas_rgs12_flux}). 
Both spectra show continua consistent with a 
multi-component thermal model that includes a number of Lyman\,$\alpha$ 
lines of hydrogen-like ions, helium-like ions, as well as a few Lyman\,$\beta$ 
lines. Fluorescent lines from lower ion stages of Fe and Si are also 
visible. The high energies are dominated by the presence of the 
``Fe\,K complex" (Fe\,XXVI\,Ly$\alpha$, Fe\,XXV, and fluorescent Fe\,K lines) 
in the range  1.7--2\,\AA. The two spectra 
of this aggregate look similar, except that the fluorescence feature is 
weaker in the XMM-{\it Newton} spectrum. 
However, our RGS spectrum reveals many more details than the {\it Chandra}
one because of the XMM-{\it Newton}'s larger effective aperture as well
as an extension to longer wavelengths and
because in 2004 the soft X-rays of \gcas\ were attenuated far less than
in 2001. These circumstances allowed us to probe to lower 
temperatures and to better judge the number of components required 
to fit both lines and continua in the soft X-ray region.

\subsection{Selection of multi-component modeling}

Our models are the results of fitting both the continua and the line
strengths in a multivariable solution with either the {\it mekal} 
formulation for optically thin plasmas, or a related model,
{\it vmekal}, in 
which individual elemental abundances can be determined independently. 
A {\it Gaussian} line was included to describe the fluorescence 
feature arising from low-ion stages of Fe at 1.94\,\AA\ (6.4\,keV), 
not incorporated in the {\it mekal} code. In all cases we adopted the {\it phabs} 
code to describe the photoelectric absorption.  When required, we assumed 
the Hipparcos distance of 188\,pc for \gcas\ \citep{Perryman97}.

\begin{figure}
 \centering
\includegraphics[angle=-90,width=9cm]{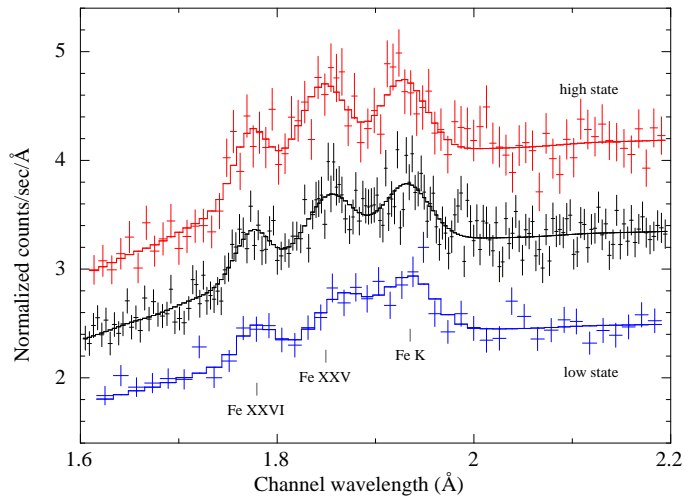}
 \caption{The Fe\,K$\alpha$ complex and its best fit determined from the model 
bremsstrahlung + 3 Gaussian lines for the low and high-state spectra (bottom and top, respectively) and the integrated spectrum. See details in Section \ref{sct:fekcomplex}.} 
\label{fig:fekcomplex}
 \end{figure}

   The high quality of our EPIC $pn$ and RGS spectra allowed us to determine
in detail the thermal components required to fit the line and continuum
fluxes. This was especially important in allowing us to test whether the
plasma has a continuous range of temperatures. Initial explorations 
showed that contributions from 3 or 4 temperatures are
necessary. In addition, otherwise mediocre fits to the continua (especially
the soft continuum) required us to consider a two-absorption column model.  
We found first the same hot temperature that many other investigators have 
reported. 
 We will refer to this component as $k$T$_Q$. 
 This component has a
  12\,keV temperature and dominates the total X-ray flux.
The carbon and nitrogen line spectra are emitted by a cool plasma
  with a temperature $k$T$_1$ of $\sim$ 0.1\,keV.
  The unabsorbed luminosity of
  this component, integrated over the range 0.2--12\,keV, is
  $\sim$ 6.8--8$\times$10$^{31}$\,erg\,s$^{-1}$.
  One or two intermediate (warm) temperatures are likewise required
  to explain the presence of lines of oxygen, neon, magnesium, {\it silicon,}
  a few Fe ions with L-shell configurations
   (Fe\,XVII-XXIV), and the Fe\,XXV line
  (Figures \ref{fig:spctgcas_rgs12_flux} and \ref{fig:fekcomplex}).

\begin{figure} \centering{ 
\includegraphics[bb=0.6cm 5.2cm 20.5cm 24.8cm,clip=true,width=9cm]{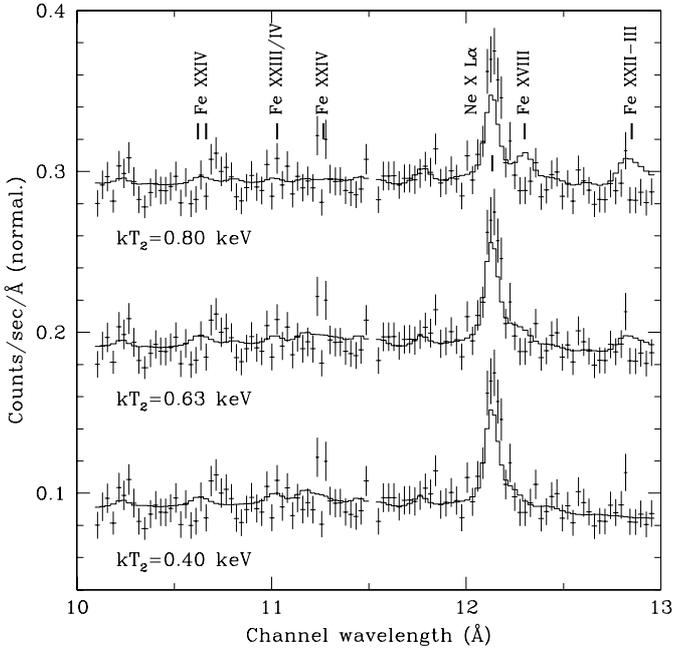}}
\caption{The observed RGS2 spectrum in the 10--13\AA\ region for
three trial values of $k$T$_2$ bracketing our
model M2, in each case offset vertically for convenience. 
This figure depicts the Ne\,X\,Ly$\alpha$ and predicted 
positions of Fe lines arising from several ions, 
and the preferred temperature $k$T = 0.63\,keV for $k$T$_2$. 
These predictions allow us
to place limits on the lower limit to the range of
the warm component, $k$T$_2$ (see text). The emission bump at
10.23\,\AA\ arises from Ne\,X Ly\,$\beta$, and it is not fit
in the range of temperatures shown.
 \label{fig:neferegion}}
\end{figure}

Before refining the initial 3 and 4 component models further, we 
attempted to determine whether these multiple thermal components were 
discrete or could be part of a continuous distribution, as might 
be expected in a single, thermally differentiated plasma.  
To continue tests using this initial set of simple models, we froze
all but one temperature in our 4-T model and computed models for 
the remaining temperature specified over graduatedsteps. 
We made movies of these results overplotted with the observations 
and determined those values of the scanned temperature which allowed us
to judge the best agreement with the observations. This technique also 
allowed us to note whether predicted lines are not observed as well as
to test whether the principal plasma components are essentially monothermal.
Figure\,\ref{fig:neferegion} demonstrates
this for $k$T$_2$ in M2 (4-T), which is necessarily confined within the
range 0.5 and 1\,keV. This plot shows the absence of Fe\,XXIII and strong Fe\,XXIV
lines, as marked on our 10--13\AA\ segment
of our RGS2 spectrum
(RGS1 shows a gap at this wavelength). These lines would be stronger
visible if the temperature were in the range
0.7--0.9\,keV, and yet they are not seen. This fact eliminates the
possibility that $k$T$_2$ is a distribution of temperatures extending to 
higher temperatures than  0.7\,keV.  On the lower bound our models 
predict the presence of the Fe\,XVIII 14.2\,\AA\ line for 
temperatures $\le$0.5\,keV, and this is not visible either. In addition,
the abrupt disappearance of components from the O\,VII complex 
in the models shows that the absence of intermediate temperature plasma 
extends from 0.5\,keV to about 0.2\,keV. Altogether
these diagnostics indicate that the $k$T$_{2}$ component in 
the \gcas\ environment is limited to a very narrow range
of temperatures centered near 0.6\,keV and that it is distinct from a 
lower or higher plasma component. This statement can probably 
be extended to the spatial separation of the warm and cool plasmas as well.

  Proceeding with this analysis,  
the weak Mg\,XII Ly\,$\alpha$ line hints the existence of 
a warm $k$T$_3$ component at $\sim$ 2\,keV, although
some of our models have managed to achieve fits without it. The absence
of yet other Fe L lines indicates that there is likely no plasma emission 
in the range of 2.5 to at least 5\,keV. Also, the presence of the 
Fe\,XXVI Ly$\alpha$ feature, the main {\it line} diagnostic in the 
hard X-ray spectrum of \gcas, requires a temperature of at least 8\,keV.
The Fe\,XXV Ly$\alpha$ line requires a value significantly lower than 
this limit, and this means that it cannot be formed in the $k$T$_Q$ plasma 
component alone.  In Figure\,\ref{fig:neferegion} a feature due to
 Ne\,X Lyman\,$\beta$ is present. This feature, visible in both the
individual RGS1 and RGS2 spectra,
cannot be easily fit with our thermal equilibrium ({\it mekal})
models. 

 It is quite possible that the $k$T$_Q$ component consists of 
some distribution of temperatures around a mean value of 12--13\,keV. 
This inference is supported by the
occasional color variations in the high energy light curve
of \gcas~ (SRC98), suggesting the presence of hot many separated exploding 
gas volumes having a small range of temperatures. 

  The upshot of these considerations the presence and absence of various 
lines in this spectrum forces the conclusion that the plasma does not exhibit 
a continuous Differential Emission Measure (DEM). Rather, there are gaps 
over the ranges of about 2.5--8\,keV, 0.7--1\,keV, and 0.2--0.5\,keV.
S04 had suspected that a continuous DEM was inappropriate 
but could not state it conclusively because of the pronounced
photoelectric absorption of the {\it Chandra} soft-energy spectrum. 

\begin{table*}[t!] \caption{Spectral fit results using the 3-T and 4-T models.\label{tbl:gcas_4T}}   
\centering           \begin{tabular}{r c c c c }   
\\[-0.5ex]        
\hline                    
\hline\\[-2.2ex]
&\multicolumn{2}{c}{hot component, 1-col.} && hot component, 2-col.\\
\cline{2-3} \cline{5-5}\\[-2.2ex]
& 3-T & 4-T && 3-T\\ 
& (M1) & (M2) && (M3)\\ 
\hline\\[-2.2ex]

N$_\mathrm{H_{a}}$ (10$^{22}$ cm$^{-2}$)    	 	&	0.22$^{+0.04}_{-0.03}$		 & 0.24$^{+0.04}_{-0.04}$	&&	0.24$^{+0.04}_{-0.04}$ \\[0.8ex]
$k$T$_{1}$ (keV)				    	&	0.10$^{+0.01}_{-0.01}$		 & 0.11$^{+0.01}_{-0.01}$	 &&	0.11$^{+0.01}_{-0.01}$\\[0.2ex]
$f_{{\rm T}_{1}}$ (erg cm$^{-2}$ s$^{-1}$)	    	&	$\sim$ 1.6$\times$10$^{-11}$ (7.3\%)	 & $\sim$ 1.9$\times$10$^{-11}$ (8.4\%)  &&		$\sim$ 1.8$\times$10$^{-11}$ (8.0\%)\\[0.2ex]
EM$_{\rm T_{1}}$ (10$^{55}$ cm$^{-3}$)	    	 	&	$\sim$ 0.5 (11.5\%)			 & $\sim$ 0.7 (15.4\%)	&& $\sim$ 0.6 (12.9\%)		 \\[0.8ex]
$k$T$_{2}$ (keV)				    	&	0.64$^{+0.03}_{-0.04}$		 & 0.64$^{+0.03}_{-0.03}$	 &&	0.62$^{+0.03}_{-0.03}$\\[0.2ex]
$f_{{\rm T}_{2}}$ (erg cm$^{-2}$ s$^{-1}$)	    	&	$\sim$ 4.0$\times$10$^{-12}$ (1.8\%)	 & $\sim$ 4.2$\times$10$^{-12}$ (1.9\%) &&		$\sim$ 4.0$\times$10$^{-12}$ (1.8\%)\\[0.2ex]  
EM$_{\rm T_{2}}$ (10$^{55}$ cm$^{-3}$)	    	 	&	$\sim$ 0.05 (1.1\%)			 & $\sim$ 0.06 (1.3\%)  && $\sim$ 0.06 (1.3\%)		 \\[0.8ex]
$k$T$_{3}$ (keV)				    	&	...				 & 2.40$^{+0.38}_{-0.26}$	 &&	...\\[0.2ex]
$f_{{\rm T}_{3}}$ (erg cm$^{-2}$ s$^{-1}$)		&	...				 & $\sim$ 1.2$\times$10$^{-11}$ (5.3\%)  &&	...  \\[0.2ex]
EM$_{\rm T_{3}}$ (10$^{55}$ cm$^{-3}$)	 		&	...				 & $\sim$ 0.2 (4.4\%)	&&	... \\[0.8ex]

$k$T$_{Q'}$ (keV)				    	&	...				 & ...	 &&	12.38$^{+0.30}_{-0.28}$ $^{(*)}$\\[0.2ex]
$f_{{\rm T}_{Q'}}$ (erg cm$^{-2}$ s$^{-1}$)		&	...				 & ...  &&	$\sim$ 3.2$\times$10$^{-11}$ (14.3\%)\\[0.2ex]  
EM$_{{\rm T}_{Q'}}$ (10$^{55}$ cm$^{-3}$)	 		&	...				 & ...	&&	$\sim$ 0.6 (12.9\%) \\[0.8ex]

N$_\mathrm{H_{b}}$ (10$^{22}$ cm$^{-2}$)    	 	&	0.029$^{+0.002}_{-0.002}$	 & 0.022$^{+0.002}_{-0.002}$	 &&	0.019$^{+0.003}_{-0.003}$\\[0.8ex]
$k$T$_{Q}$ (keV)				    	&	12.55$^{+0.31}_{-0.29}$		 & 14.32$^{+0.59}_{-0.55}$	 &&	12.38 $^{(*)}$ \\[0.2ex]
$f_{{\rm T}_{Q}}$ (erg cm$^{-2}$ s$^{-1}$)		&	$\sim$ 2.0$\times$10$^{-10}$ (90.9\%)	 & $\sim$ 1.9$\times$10$^{-10}$ (84.4\%)  &&	$\sim$ 1.7$\times$10$^{-10}$ (75.9\%)\\[0.2ex]  
EM$_{{\rm T}_{Q}}$ (10$^{55}$ cm$^{-3}$)	 		&	$\sim$ 3.8 (87.4\%)			 & $\sim$ 3.6 (78.9\%)	&& $\sim$ 3.4 (72.9\%)		 \\[0.2ex]

$Z_{{\rm T}_{Q,Q'}}$ ($Z_{\odot}$)		    	&	0.13$^{+0.01}_{-0.01}$		 & 0.10$^{+0.02}_{-0.02}$	&&	0.12$^{+0.01}_{-0.01}$ \\[0.8ex]
												 				 
Line (keV)				     		&	6.4$^b$				 & 6.4$^b$			 &&	6.4$^b$\\[0.2ex]
$\sigma_{\rm Line}$ (keV) 		  		&	0.01		 & 0.01	 &&	0.09$^{+0.04}_{-0.03}$\\[0.8ex]
												 				 				 
$f_{tot}$ (erg cm$^{-2}$ s$^{-1}$) 			&	$\sim$ 2.2$\times$10$^{-10}$	 & $\sim$ 2.2$\times$10$^{-10}$  &&	$\sim$ 2.3$\times$10$^{-10}$\\[0.2ex]  
EM$_{tot}$ (10$^{55}$ cm$^{-3}$)                        & $\sim$ 4.4	& $\sim$ 4.6	&& $\sim$ 4.7 \\[0.8ex]
$\chi^{2}_{\nu}$/d.o.f.$^{\mathrm{a}}$	     		&	1.47/1532			 & 1.44/1529	&&	1.46/1530\\[0.2ex]
\hline
                                  \end{tabular} 
\begin{list}{}{}
 
\item $^{\mathrm{a}}$ degrees of freedom; $^{\mathrm{b}}$ frozen parameter.
\item Notes: Fluxes are given unabsorbed in the 0.2--12\,keV energy band. 
In parentesis are the fluxes and EM in \% of the total values. 
Plasmas with solar abundances, except for the [Fe] of the hot component. 
Quoted errors are at the 90\% confidence level.
Base model: 
\textsc{N$_{\rm H_a}*$(T$_1$+T$_2$+T$_3$+T$_{Q'}$)+N$_{\rm H_b}*$(T$_Q$)}; see Section \ref{sect:modelfit} for details.

\end{list}
\end{table*}

\subsection{Model fitting}
\label{sect:modelfit}

  The above procedures allowed us to refine our initial 3-T and 4-T
models, utilizing both line and continuum flux information. Further
attempts to search for a continuous Differential Emission Measure
from the continuum alone, such as with a cooling flow
{\it cemekl} led to significantly degraded fits even in the 
continuum.\footnote{The {\it cemekl/cevmkl} code describes a multi-temperature 
plasma based on the {\it mekal} code, in which the emission measures 
of the plasma scale with their temperature as (T/T$_{\rm max}$)$^{\alpha}$;
$\alpha$ = 1 corresponds to the adiabatic case.}

The fits converged for $k$T$_{\rm max}$ $\sim$ 42\,keV and $\chi^{2}_{\nu}$ = 
2.4.  Freezing $k$T$_{\rm max}$ to 12\,keV produced an unacceptably high
$\chi^{2}_{\nu}$ = 6.4. We next tried a composite
model {\it cemekl + mekal}. This resulted in temperatures of 
$k$T$_{\rm max}$ $\sim$ 12\,keV and $k$T $\sim$ 0.9\,keV
for each component, respectively, and a $\chi^{2}_{\nu}$ = 1.75. This
model failed to describe the continuum at high energies and underpredicted
the following features: the Fe\,XXV line in the Fe\,K$\alpha$ complex,
the Fe L-shell lines, OV\,III\,Ly$\beta$, the $fir$ triplet of O\,VII, 
N\,VII\,Ly$\alpha$, and Ne\,X\,Ly$\alpha$.
Allowing the {\it cemekal} $\alpha$ parameter to vary did not improve the fits.

\begin{figure*} \centering{ 
\includegraphics[bb=1.7cm 11.5cm 19.9cm 24.4cm,clip=true,width=17.5cm]{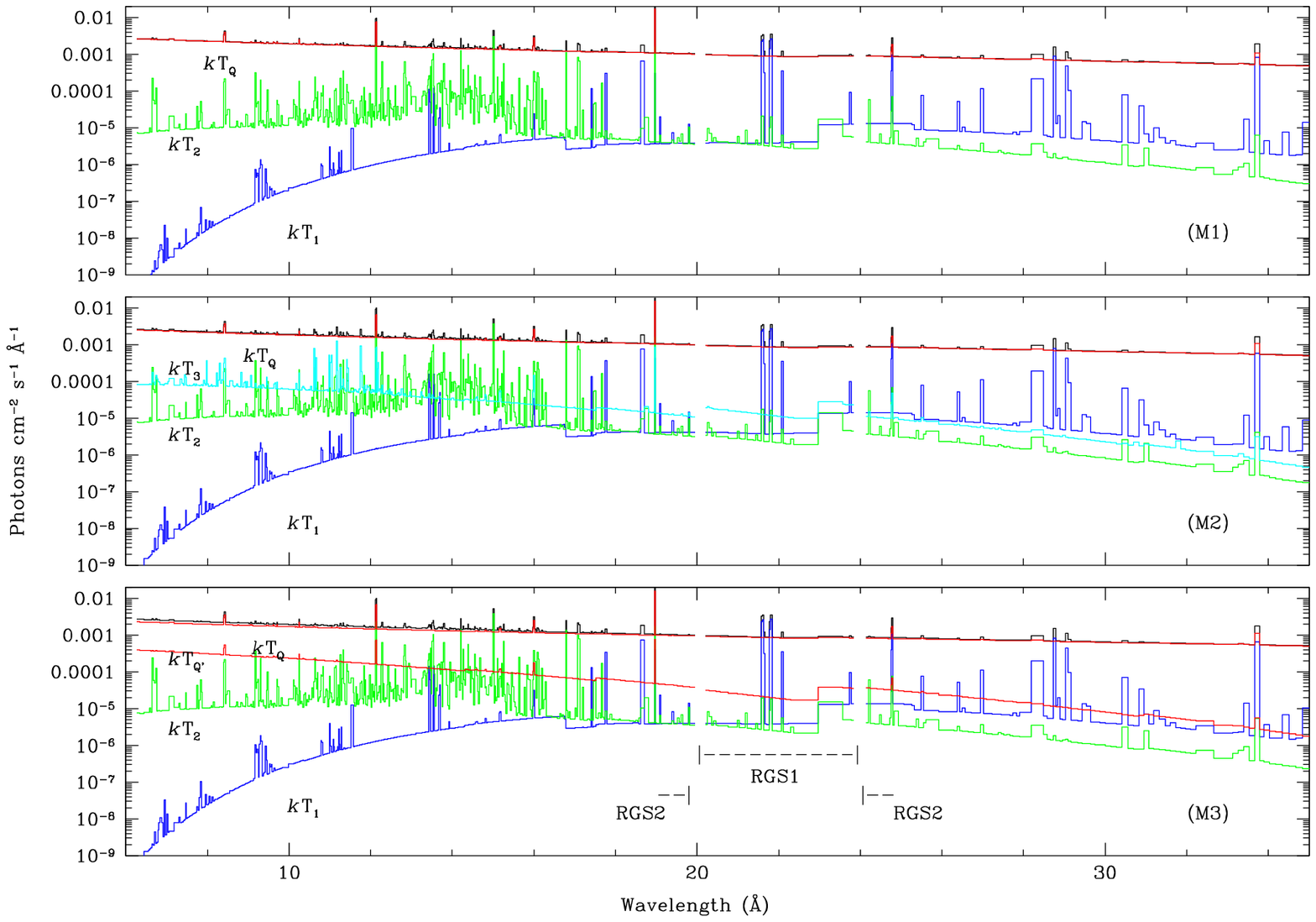}
\caption{Unfolded spectrum from RGS1 and RGS2 
for each model in Table \ref{tbl:gcas_4T}.
The blue, green, cyan (light blue) and red lines correspond to the 
$k$T$_{1}$, $k$T$_{2}$, $k$T$_{3}$ and $k$T$_{Q}$ (and $k$T$_{Q'}$), 
respectively, while the black lines correspond to the composite model.
}
\label{fig:unfoldspctnew}} \end{figure*}

\begin{figure} \centering{ 
 \includegraphics[bb=1.7cm 15.3cm 10cm 24.4cm,clip=true,width=8cm]{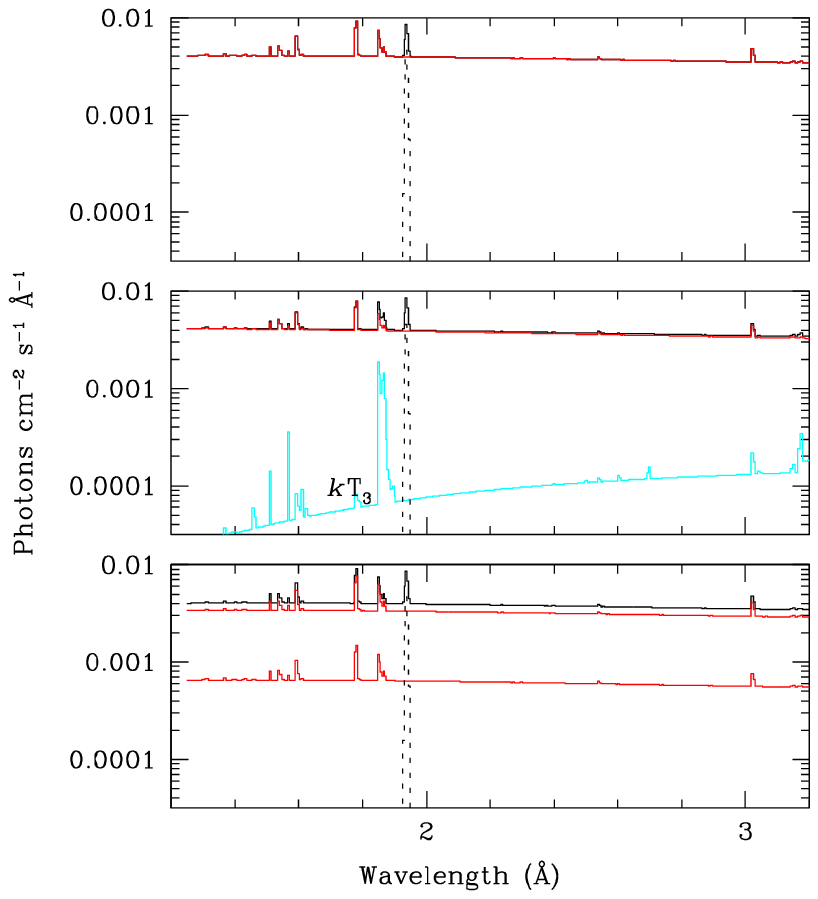}
\caption{Unfolded spectrum from EPIC $pn$, centered at the 
Fe\,K$\alpha$ complex for each model in Table \ref{tbl:gcas_4T} (see Fig. \ref{fig:unfoldspctnew} for details).
The dotted lines refer to the {\it Gaussian} line at 1.94\,\AA (6.4\,keV).
}
\label{fig:unfoldspctpn}} \end{figure}

Table \ref{tbl:gcas_4T} shows the results of our analysis in terms 
of three basic models. Columns 2, 3, and 4 list the values determined 
for the temperature, column density, the modeled unattenuated flux, 
the resulting percentage of the total X-ray flux, and
emission measures for the three and four component models,
and the Fe abundance derived from the Fe\,XXV and Fe\,XXVI\,Ly$\alpha$ ions.
In all cases the high temperature component we designate as
$k$T$_Q$ was found to lie near 12--14\,keV.
Column 2 of Table \ref{tbl:gcas_4T} shows the solution for a 3-T model,
``(M1)," that includes a cool and warm components ($k$T$_1$ and 
$k$T$_2$, 
respectively) as well as 
the dominant one, $k$T$_Q$. The cool/warm component fluxes were 
attenuated by a column N$_{\rm H_a}$ while the hot one was attenuated by 
N$_{\rm H_b}$. Our 4-T model (column 3 of Table\,\ref{tbl:gcas_4T}; ``M2") gives 
the solution for the addition of a warm component having $k$T$_{3}$ 
and also attenuated by N$_{\rm H_a}$. The final column of the table, ``(M3)," is 
the 3-T solution again. However, this time the hot component was attenuated 
by both N$_{\rm H_a}$ and N$_{\rm H_b}$ columns. In this model, the dominant hot 
subcomponent ($k$T$_Q$), 
affected by the individual N$_{\rm H_b}$ column, contributes 
to $\sim$ 76\% of the total flux. 
The second hot subcomponent 
 ($k$T$_Q'$), warm ($k$T$_2$), and cool ($k$T$_1$) 
components were affected by a common N$_{\rm H_a}$ absorption column. 
 Figures \ref{fig:unfoldspctnew} and \ref{fig:unfoldspctpn} show the unfolded spectrum for each model,
allowing the reader to judge which of the various observed lines is formed 
in which plasma component. In our solutions we found that
N$_{\rm H_a}$ $\sim$ 2$\times$10$^{21}$ cm$^{-2}$ and N$_{\rm H_b}$ $\sim$ 
2--3$\times$10$^{20}$ cm$^{-2}$.  By comparison,
the UV and H$\alpha$ determined ISM column density 
to \gcas\ from the literature is a scant 1--2$\times$10$^{20}$ cm$^{-2}$,
which leaves only a little room for absorption of the soft X-rays within
the source.
 
The 4-T fit we found included components $k$T$_1$,  $k$T$_2$, $k$T$_3$, and
$k$T$_Q$ having values of about 14\,keV, 2.4\,keV, 0.6\,keV, and
0.1\,keV, respectively.  For the 3-T models, the $k$T$_3$ component was omitted,
but this omission must be compensated for in the Fe\,XXVI\,Ly$\alpha$ strength by
decreasing the temperature from 14 to 12.5\,keV.

It is worth noting that any of our models give us the desired $\chi^{2}_{\nu}$ $\sim$ 1.
The residuals are mostly in the lines, not in the continuum, and the relatively
high -- although even acceptable -- values of $\chi^{2}_{\nu}$ 
(Table \ref{tbl:gcas_4T}) 
are due in part to an inadequate description of the individual line
profiles.

A key point to take away from the Table \ref{tbl:gcas_4T} is that although
84--90\% of the total flux is radiated by the $kT_Q$ (or $kT_Q$+$kT_Q'$) component(s), 
almost all the lines except for the Fe\,K complex is contributed by the cooler
$k$T$_{1-3}$.  The hot $kT_Q$ component therefore provides no kinematic information. 
However, the possibility of multiple absorption columns and the presence of 
the fluorescence features of Fe and Si do provide potential geometrical
descriptions for the emitting volumes. 

   Although S04's solution for the 2001 spectrum was similar in that it
required a 4 thermal component fit, there were some small differences in
detail that are significantly different. One of the clearest differences
is with the precise temperature of a warm component (their 
``$k$T$_3$" $\sim$ 0.4\,keV cf. the $k$T$_2$ $\sim$ 0.6\,keV herein). 
A small increase in the O\,VIII/O\,VII Lyman\,$\alpha$ line ratio for
the XMM-{\it Newton} spectrum is indicative of the higher value we find. 
If a warm temperature 
is a quasi-permanent feature of the X-ray spectrum, it has definitely 
shifted from 0.4\,keV to 0.6\,keV between 2001 and 2004. The hot
($\sim$ 12--14\,keV) and cool ($\sim$ 0.1\,keV) component temperatures
overlap. Although we cannot be sure if  $\sim$ 2\,keV plasma
existed in 2004, its possible presence is consistent with emission
powered by a similar thermal component in the 2001 spectrum in their
2001 {\it Chandra} observation.

  As already noted, one of most important differences from
earlier results is that no high absorption column was needed
to fit this spectrum.  
S04 found that the {\it Chandra} spectrum
required a fitting with two hot plasma absorption columns. 
The first  contributed 10--30\% of the 12\,keV component and was
attenuated by an high absorption column of 10$^{23}$ 
cm$^{-2}$. The second component, which is the dominant flux
contributor to the observed spectrum at high energies, 
was attenuated by a column of $\sim$ 3$\times$10$^{21}$ cm$^{-2}$. 
In contrast, most of
the X-ray flux observed by XMM-{\it Newton}
is affected by one dominant absorption column equivalent 
to N$_{\rm H}$ $\sim$ 2--3$\times$10$^{20}$ cm$^{-2}$, as derived from our
3-T and 4-T model with a single absorption column. However, for our
models we were only able to fit the continua of our 
{\it combined} EPIC $pn$ and RGS spectra with two absorption columns. 
Although the absorption of the N$_{\rm H_b}$ column, at
$\sim$ 2--3$\times$10$^{20}$ cm$^{-2}$, is very low, it has
much influence because it affects most (62--96\%) of 
the total emission (see Table \ref{tbl:gcas_4T}). The other absorption 
column has only a minor influence in absorbing the flux even though it is 
higher by a factor of 7--10 (N$_{\rm H_a}$ $\sim$ 2$\times$10$^{21}$ cm$^{-2}$). 
We note that the temperatures derived for each component in the models 
with one, as opposed to two, absorption columns are consistent with one
another.
The two column model also improved the descriptions of the lines 
formed at low energies, such as
Ne\,X\,Ly$\alpha$, and O\,VIII\,Ly$\alpha$.
The improvement of the fit with the second column over
a model with a single column
is supported by the F-test statistic, stating 
with a probability of only 4.7$\times$10$^{-5}$ that the derived 
decrease in $\chi^{2}_{\nu}$ 
associated with the second column column is due to chance. 
The important conclusion from this is that the absence of a high absorption 
column as derived from the XMM-{\it Newton} observation suggests a 
dramatic rearrangement of cold circumstellar gas in the vicinity of 
the X-ray sources between 2001 August and 2004 February.  
Contrary to what is arguable for the 2001 epoch, it cannot be said 
that in 2004 the circumstellar disk was in front of a significant fraction 
of the soft X-ray source(s).

In order to estimate the origin of the long-wavelength lines in our models,
we constructed a lineless continuum by excluding of the spectral regions lines
in the 10--17\AA~ range and fit the resulting pure continuum spectrum to a
bremsstrahlung model. This model gave a value a slightly hotter temperature
13.2 keV for our hot component with a $\chi^{2}_{\nu}$ of 1.17. 
The addition of a second bremsstrahlung component did not improve the fits 
($k$T $\sim$ 0.04 keV; $\chi^{2}_{\nu}$ of 1.17).
This result suggests that the determination of the 
low component temperatures, $k$T$_1$ and $k$T$_2$, is strongly driven
by the lines alone.

Although all of our discussion has taken place in the context
of equilibrium models, we agree with Smith et al. (2004), who suggested that
the warm and/or cool plasma components could also result from nonequilibrium
ionization processes, e.g. from the sudden impact of gas parcels ejected
by flares into a dense stationary medium such as the Be disk. 
We suspect this could also be the reason why we observe Lyman\,$\beta$
transitions, 
such as the Ly$\,\beta$ transitions of Ne\,X and O\,VIII,
 that are not easily fit with equilibrium {\it mekal} models.
A similar explanation might be also important in understanding the
Fe\,XXVI Ly$\beta$ line at 8.3\,keV of HD\,110432, 
reported by \citet{Lopes07}, 
and which could not be fit with standard models for the ranges of
temperature considered for this feature. 
The strong Lyman\,$\beta$ features hint at excitation temperatures in
these plasmas that are higher than the ionization temperatures we
found in Table \ref{tbl:gcas_4T}.

\subsection{Abundances and anomalous line strengths} 

 For the most part the line strengths are consistent
with solar abundances in our 3-T and 4-T models. 
The first and most conspicuous exception to this general result is the 
finding from Table\,\ref{tbl:gcas_4T} that the Fe abundance from K-electron
Fe ions is 0.12${\pm 0.02}$\,Z$_{\odot}$. This is a factor 
of two lower than any other studies in the literature (the S04 result was 
0.24$\pm{0.02}$\,Z$_{\odot}$). 
Because K-shell ion abundances of 
0.1\,Z$_{\odot}$ and 0.24\,Z$_{\odot}$ are significantly 
different from one another, our result clarifies in a new way that the Fe 
abundance derived from K-shell ions changes with time. 
Thus, the K-shell result may not reflect a global elemental abundance.

As also found by S04, the Fe abundance from L-shell ions is significantly 
higher than the Fe abundance from K-shell ions. 
We pursued the investigation of the Fe abundance from the Fe L-shell 
lines with the models M1, M2 and M3. In a new set of models we fixed 
other variables and solved for the Fe abundance of the soft temperature 
components $k$T$_1$ and $k$T$_2$. 
Our models suggest that the Fe L-shell abundance could be 0.4 to 1.3$\times$Z$_{\odot}$. 
This exercise verifies the distinctness of the Fe abundance arising from the
L-shell lines on one hand and the K-shell lines on the other.
For completeness note that a similar K- and L-shell Fe anomaly was
reported by \citet{Lopes07} for \hd. 

The second apparent departure from solar abundances comes
from the enhanced strengths of the lines of hydrogen and helium-like
ions of neon and nitrogen. The formal abundances we find from our models
are Z$_{\rm N}$=3.96$^{+0.87}_{-0.69}$$\times$Z$_{\rm N,\odot}$ and
Z$_{\rm Ne}$=2.63$^{+0.27}_{-0.28}$$\times$Z$_{\rm Ne,\odot}$, assuming solar abundance for the other metals in the softest plasma and allowing for line broadening and bulk velocities. However, the allowed range of Ne abundances depends on the number and nature of the parameters left free in the fit, the most important being the assumed Fe L-shell abundance and the line broadening velocity. The hottest component dominates the NeX line flux with a smaller ($\sim$ 1/3) contribution from the ${\rm T}_{2}$ thermal component. For instance, using bvapec instead of mekal and leaving Fe L-shell abundance free yields slightly lower abundances with 
Z$_{\rm Ne}$=1.65$^{+0.38}_{-0.29}$$\times$Z$_{\rm Ne,\odot}$ and 
Z$_{\rm Fe L-shell}$=0.41$^{+0.18}_{-0.11}$$\times$Z$_{\rm Fe,\odot}$. This effect is due to the presence of a FeL-shell forest around the NeX\,L$\alpha$ line and to the fact that the continuum level depends on the description of the FeL-shell lines. 
Although the limits for the abundances are not well determined, there is a clear excess above the solar values for N and Ne. The NeX line peak velocity is in the range of -110 to +230 km\,s$^{-1}$ and displays a broadening velocity comparable to that of other bright emission lines.

A similar analysis was applied to the Chandra spectrum in order to determine whether the Ne enhancement was present or not in 2001 -- it is not possible to expand the investigation to N abundance due to the low SNR of the NVII\,L$\alpha$ line. 
The non-velocity broadened {\it mekal} model yields Z$_{\rm Ne}$ = 0.75$\pm$0.19$\times$Z$_{\rm Ne,\odot}$ consistent with the value obtained by \citet{Smith04}. However, since the observed NeX line apppears significantly wider than modeled, we used the {\it bvapec} model to take into account such a broadening. The larger NeX EW impacts the determination of the abundance yielding Z$_{\rm Ne}$=1.52$^{+0.24}_{-0.24}$$\times$Z$_{\rm Ne,\odot}$ (or Z$_{\rm Ne}$ = 1.20$^{+0.28}_{-0.24}$$\times$Z$_{\rm Ne,\odot}$ if the oxygen and iron L-shell abundances are left free, in which case we obtain Z$_{\rm Fe L-shell}$ = 0.46$^{+0.09}_{-0.09}$$\times$Z$_{\rm Fe,\odot}$). 
The broadened line profile fits the observed Ne line much better, 
with peak velocities then in the range of -60 to 260 km\,s$^{-1}$ and broadening velocities of $\sim$ 500 $\pm$ 200 km\,s$^{-1}$. These values are
consistent with those derived from the RGS spectrum.

We note carefully that whereas the fluxes in these lines scale
linearly with abundance, the equivalent widths, as formally defined with
respect to the neighboring continuum flux, scale far more slowly because
of the contribution to the local bound-free opacities from ions in a N-Ne
rich plasma. Thus, we found an increase of only 35\% with respect
to the equivalent width measured in the {\it Chandra} spectrum. 
Our models with XSPEC validate this mild increase in equivalent width
with abundance in detail. Given this 
reality, we anticipate that the true abundance errors are larger
than those XSPEC computes based on photon statistics. Nonetheless, 
the anomalous excess cannot be discounted. Moreover, the
possibility that the X-ray environment could be nitrogen-rich by a
factor of 3-4 is unremarkable because enhancements are already a hallmark 
of massive stars evolving off the main sequence, although the
reasons for this are still under discussion \citep{Hunter}.
However, the apparent neon enhancement is of much greater interest 
because in a stellar evolution context neon is produced by carbon burning 
shortly before supernova detonation or in the interiors of some white 
dwarfs, and dredging to the surface of such elements is possible by 
various mixing processes. 

  Suggested alternatives to the straightforward abundance 
interpretation of the line strengths are the following:
(i) instrumental artifact or a cosmic ray; 
(ii) an inappropriate temperature used in the modeling; and
(iii) the line is strengthened by a microturbulent like broadening.
  Each of these possibilities may be dismissed in turn. (i) is unlikely 
because separate spectrum extractions from the positive and negative
detector halves show consistent profiles.
Likewise, we may rule out (ii) because a temperature of 0.6\,keV is already 
ideal for Ne\,X formation, and a stronger feature cannot be produced
given a standard abundance. Possiblity (iii) would require a large optical
depth and strengthening of the line. However, even though
the column density is probably high enough for photons to experience
more than one mean free path through their
transits across the medium, they are nonetheless scattered 
coherently. Therefore the line widths are unaltered by these histories.
With these possibilities ruled out, we are forced to
conclude that the neon and nitrogen abundances in the X-ray plasmas
of \gcas\ are high. 

   In considering the anomalous Fe abundance derived from
Fe-K lines, but normal abundance from Fe-L lines in the \gcas\ spectrum, 
S04 suggested that this 
could arise from an inverse FIP (first ionization potential) effect
in the \gcas\ environment that is similar to that found
in the coronae of the Sun and other magnetically
and X-ray active cool stars, such as AB Dor \citep{Gudel}.
Although the details are still unclear, it appears that in a magnetic
plasma low density environment differential ponderomotive forces, produced
by wave heating and dependent on their first ionization potential, can 
prevent certain ions from migrating across a plasma and resulting in an
altered measured chemical abundance \citep{Laming}. 
Whether such a process is actually active
in the environment of \gcas, let alone whether it extends to
Ne$^{9+}$ ions, must be considered speculative. Its attractiveness lies
in its potential to explain the different Fe abundances derived from
K-shell and L-shell ions.
Similarly, now in the framework of the accretion model, it is not clear, first,
how Ne could be preferentially ejected from the white dwarf atmosphere or,
second, how Ne could be enhanced without enhancing at the same time
C, O, and Mg abundances.

\subsection{The Fe\,K$\alpha$ complex}
\label{sct:fekcomplex}

In evaluating the equivalent widths (EW) of each emission iron line of the
Fe\,K complex, we used the 5--10\,keV photons acquired during low background 
phases of the XMM-{\it Newton} satellite's orbit. 
Our approach was to limit consideration of spectra accumulated only
during low (L) and high (H) flux states observed in \gcas\, 
(defined as the lowest and highest one-third fluxes of the total
distribution in our light curve) and also
during some 1000 time windows in which the detector background was low (see Fig. \ref{fig:lc_hrd_gcas}-b).
Finally,
we applied an absorbed bremsstrahlung model in order to describe the 
underlying continuum and three {\it Gaussian} lines to account the iron lines. 
 Table \ref{tbl:EW_FeKa} and Fig. \ref{fig:fekcomplex} show the results.  The measured centroid energy of 
each modeled {\it Gaussian} line is, within the errors, in 
agreement with the theoretical values for the fluorescent, helium-, and 
hydrogen-like components of the Fe\,K$\alpha$ complex. 
It is possible that there may be a slight inverse sensitivity of the
fluorescence strength (relative to the Fe\,XXVI line), though this
inference is of marginal statistical significance.

\begin{table} \caption{Parameters of the emission lines of the Fe\,K$\alpha$ complex from a bremsstrahlung + 3 Gaussian lines model.}              
\label{tbl:EW_FeKa}        \centering           \begin{tabular}{l c c c c c c c c c c c c}            
\hline                    
\hline\\[-2.2ex]

 & $\lambda_{\rm C}$ & EW     & Flux$^{\mathrm{a}}$  \\ 
 & (\AA)             & (m\AA) & ($\times$10$^{-5}$) \\ 

\hline\\[-2.2ex]
all observation: \\
Fe\,K (fluorescence)	& 1.9312[60] & 10[2] & 4.4$^{+0.8}_{-1.0}$ \\	   [0.2ex] 
Fe\,XXV 		& 1.8533[55] & 12[3] & 5.3$^{+1.4}_{-0.9}$ \\	   [0.2ex]	     
Fe\,XXVI Ly$\alpha$   	& 1.7737[51] &  7[2] & 3.1$^{+0.7}_{-0.5}$ \\	   [0.2ex] 
\hline\\[-2.2ex]
low state: \\
Fe\,K (fluorescence)	& 1.937[12]  & 13[5] & 4.1$^{+1.0}_{-1.7}$ \\	  [0.2ex] 
Fe\,XXV 		& 1.867[17]  & 12[4] & 4.1$^{+1.5}_{-1.2}$ \\	  [0.2ex]	    
Fe\,XXVI Ly$\alpha$   	& 1.781[10]  &  7[3] & 2.2$^{+0.9}_{-0.8}$ \\	  [0.2ex] 
\hline\\[-2.2ex]
high state: \\
Fe\,K (fluorescence)	& 1.9282[60] & 10[3] & 5.6$^{+1.0}_{-1.6}$ \\	   [0.2ex] 
Fe\,XXV 		& 1.8478[83] & 13[4] & 7.6$^{+2.2}_{-1.4}$ \\	   [0.2ex]	     
Fe\,XXVI Ly$\alpha$   	& 1.7737[76] &  8[2] & 4.4$^{+1.3}_{-0.9}$ \\	   [0.2ex] 

\hline
                              \end{tabular} 
\begin{list}{}{}

\item [$^{\mathrm{a}}$] Total flux in line, in units of photons\,cm$^{-2}$\,s$^{-1}$. Notes: Quoted errors are at 90\% confidence level. The temperature of the bremsstrahlung model at 1.2--4.1 \AA\ (3--10 keV) converges to $k$T = 13.36(+0.21/-0.25) keV.

\end{list}
\end{table}

In our EPIC $pn$ spectrum the
strength of the fluorescence feature (EW $\sim$ -10\,m\AA) is weaker than when Chandra observed it \citep[EW $\sim$ -19\,m\AA;][]{Smith04},
in line with the fact that the attenuation of 
soft X-rays by cold matter is less.
This suggests that the fluorescence
emission feature is formed at least partially by
the same medium that aborbs the soft-X ray flux.

Curiously, the analysis in the low and high-state spectra reveals
that the FeK fluorescence feature may be marginally
stronger in the low flux case.
Although such investigation could not be expanded to a quantitative analysis 
of all spectral range because the relatively limited signal-to-noise in the 
final spectra for RGS1/2, we suspect that there is weak 
evidence for a weakening of the SiK feature for the low flux spectrum.

\subsection{Helium-like diagnostics of electron densities}

Our spectrum covers the regions of the He-like Ne\,IX,
O\,VII and N\,VI complexes -- but the N\,VI complex falls partially 
onto gaps in the CCD for RGS1 and RGS2. 
Each of these is comprised of a so-called
$fir$ (forbidden/intercombination/resonance) line triplet 
(see Fig. \ref{fig:spctgcas_rgs12_flux}), 
and the ratio of their intensities can indicate
whether the dominant excitation process for producing this triplet
is collisional or photoionization. For example, if collisions are
dominant, the ratio $G = (i + f)/r$ $\sim$ 1 obtains, whereas 
if photoionizations dominate the exitation $G$ $\sim$ 4 
\citep[e.g.][]{Porquet00,Porquet01}.
 From the O\,VII $fir$ complex in our XMM-{\it Newton} spectrum,
we estimate G $\sim$  0.9$\pm{0.1}$ for $\gamma$\,Cas. 
The N\,VI $rif$ components are poorly measured because 
they fall within the gaps of the RGS2 detector. Nonetheless,
by computing their ratios we find nearly the same value, 
G $\sim$ 0.7$\pm{0.3}$. This suggests that the plasma 
is in the classical domain of collisional dominance.
As it happens, the $fir$
ratio alone cannot distinguish between quenching of the
forbidden transition by collisions or by photoexcitations by 
a nearby strong UV source, such as the Be star.

\begin{table*} \caption{Parameters of the strongest emission lines.}              
\label{tbl:plines}        \centering           \begin{tabular}{l c c c c c c c c c c c c}            
\hline                    
\hline\\[-2.2ex]

 & $\lambda_{\rm C}$ & EW   & \multicolumn{2}{c}{Flux$^{\mathrm{a}}$}\\ 
 & (\AA)       & (m\AA) & \multicolumn{2}{c}{($\times$10$^{-4}$)}\\ 

\hline\\[-2.2ex]

Ne\,X\,Ly$\alpha$$^c$       &  12.131[11]   &  125[18]  & 2.15[31] & (1.26$\pm$0.16) \\ [0.2ex] 
O\,VIII\,Ly$\beta$$^{c}$    &  15.997[41]   &   40[14]  & 0.55[20] & (0.31$\pm$0.16) \\ [0.2ex]
O\,VIII\,Ly$\alpha$$^{c}$   &  18.9796[49]  &  360[28]  & 3.93[31] & (1.71$\pm$0.26) \\ [0.2ex]
O\,VII\,$r$$^b$             &  21.60[90]    &  163[19]  & 1.72[20] & (0.77$\pm$0.28) \\ [0.2ex]
O\,VII\,$i$$^b$             &  21.78[92]    &  149[19]  & 1.58[20] & (0.79$\pm$0.40) \\ [0.2ex]
O\,VII\,$f$$^b$             &  22.097$^d$   &   $<$32   & $<$0.3   & (0.13$\pm$0.16) \\ [0.2ex]
N\,VII\,Ly$\alpha$$^c$      &  24.79[99]    &  296[17]  & 2.76[16] & (0.61$\pm$0.03) \\ [0.2ex]
N\,VI\,$r$$^c$              &  28.81[90]    &  $<$230   & $<$2     & ... \\ [0.2ex]
N\,VI\,$i$$^c$              &  29.09[95]    &  $<$169   & $<$1.4   & ... \\ [0.2ex]
N\,VI\,$f$$^c$              &  29.531$^d$   &  $<$10    & $<$ 0.08 & ... \\ [0.2ex]
C\,VI\,Ly$\alpha$$^d$       &  33.736[64]   &  201[29]  & 1.37[20] & ... \\ [0.2ex]

\hline
                              \end{tabular} 
\begin{list}{}{}

\item [$^{\mathrm{a}}$] Total flux in line, in units of photons\,cm$^{-2}$\,s$^{-1}$; in parenthesis are shown the values derived by \citet{Smith04} from Chandra. $^{\mathrm{b}}$ From RGS1.
$^{\mathrm{c}}$ From RGS2. $^{\mathrm{d}}$ Frozen parameter. Quoted errors are at 90\% confidence level.

\end{list}
\end{table*}

\subsection{Velocity broadening and shifts in warm plasma lines}

In Table \ref{tbl:plines} we list the observed 
centroid wavelength, the equivalent width and the flux of lines for those lines for which meaningful 
measurements could be made.  The best observed lines in the soft 
X-ray spectrum, Ne\,X\,Ly$\alpha$ and O\,VIII\,Ly$\alpha$ 
profiles are noticeably broadened.  
We evaluate the broadening of the lines by using the {\it bapec} model,
a velocity- and thermally-broadened emission model from collisionally-ionized
diffuse gas\footnote{http://cxc.harvard.edu/atomdb/}. 
The {\it Gaussian} sigma for velocity broadening converges to $\sim$ 400 km\,s$^{-1}$
for a model like M1, M2 and M3 in Table \ref{tbl:gcas_4T}, replacing the {\it mekal} by 
the {\it bapec} code. The line broadening was also measured on individual lines, not only on
the whole spectrum. For example, we found a velocity broadening of 300$\pm$70 km\,s$^{-1}$
for the O\,VIII line.
The RGS velocity is consistent with that reported by S04 (478$\pm$50 km\,s$^{-1}$).
We applied the same analysis to two RGS spectra of AB\,Dor acquired in 2000 and 
2006 (obsID 0126130201 and 0160363001 with up-to-date calibration) and found 
in each case that the broadening of the O\,VIII line was less than about 70 km\,s$^{-1}$ 
at the 90\% confidence level, proving that the lines of \gcas\ are indeed broadened.

In order to check for velocity shift in the lines, we run the models in 
Table \ref{tbl:gcas_4T} leaving the redshift parameter free. 
We found a overall redshift of $\sim$ 200 km\,s$^{-1}$ -- and values for the
parameters and $\chi^{2}_{\nu}$ consistent
with those shown in Table \ref{tbl:gcas_4T} -- slightly larger than the typical error 
on the absolute wavelength scale of $\sim$ 7 m\AA, or $\sim$ 140 km\,s$^{-1}$.

\subsection{Do our cool or warm plasma component arise in a radiative wind?}

  It has been believed for some time that X-ray emission in O and early B 
stars arises in sites associated with shocks distributed in the stars 
radiative winds \citep[e.g.][]{Leutenegger06,Cohen08}. 
Analyses of HETG spectra of
other stars in the same general region of the H-R Diagram,  for
example of $\tau$\,Sco (B0.2\,V), $\beta$\,Cru (B0.5\,III), and
$\zeta$\,Oph (O9.5\,V) by \citet{Cohen03}, \citet{Cohen08},
and \citet{Waldron05}, respectively, paint a complex picture.
The Cohen et al. analyses show that the winds of at least some
early B stars cannot be described by a standard picture of
wind-shocks, and indeed wind structures of at least $\tau$\,Sco
and $\zeta$\,Oph may be influenced by the effects of channeling
by magnetic fields in the regions surrounding the star.  The
temperatures of the gas are not yet known to vary but taken as 
a group range from about 3 to 25\,MK. 
We note that the X-ray emission measures associated with winds in these stars have not been found to be variable so far.

$\gamma$\,Cas has such a wind, as evidenced by the profiles of doublets
in UV resonance absorption lines of several ions, notably C\,IV, N\,V, 
and Si\,IV. These profiles exhibit Discrete Absorption features at
1000-1100 km\,s$^{-1}$ and edge velocites of -1800 km\,s$^{-1}$ 
\citep[][]{Doazan82,Kaper96,SRC98,SRH98}. 
Our analysis shows the presence of soft X-ray components whose temperatures, fluxes, and emission measures (Table \ref{tbl:gcas_4T}) are consistent with those of normal massive stars reported by \citet{Walborn2009}, for example. It is thus expected that at least part of such emission is emanating from the radiatively driven wind of \gcas. However, as can be seen in Fig. \ref{fig:unfoldspctnew}, the hottest component dominates also the soft part of the spectrum of \gcas\ and the contribution of a pure radiative wind is overshadowed by this hot component.
Establishing the presence of a luminosity change in the soft components supporting a non-wind origin is a complicated issue because of the presence of the Be disk itself (which may shield a part of the wind) and because 
the characteristics of the \gcas' spectra in 2001 and 2004 are different from each other.
It is worth noting that the X-ray 0.2--12 keV (unabsorbed) fluxes from the soft ($k$T $\sim$ 0.1 keV) component in 2001 and 2004 are consistent with each other ($<$ 1.7$\times$10$^{-11}$ erg\,s$^{-1}$ from Chandra and 1--2.7$\times$10$^{-11}$ erg\,s$^{-1}$ from XMM-{\it Newton}), while the flux from the $k$T $\sim$ 0.6 keV component has increased between 2001 and 2004 (from 0.7--2.2 to 2.5--6$\times$10$^{-12}$ erg\,s$^{-1}$). 
In conclusion, we cannot discard an additional contribution to the soft plasma in addition to that of a radiatively driven wind usually observed in massive stars.

\section{Conclusions}

   We have reported the second high dispersion analysis of the
X-ray spectrum of the X-ray anomalous B0.5e star \gcas\ obtained by the
XMM-{\it Newton} in 2004, and we find the following characteristics
at this epoch:

\noindent {\it a)} The emission is due to an optically thin, 
thermal medium comprised 
of 3 to 4 discrete components but dominated by a hot component having 
$k$T$_Q$ $\sim$ 12--14\,keV.  The temperature of at least one 
component ($k$T$_2$ $\sim$ 
0.6\,keV) has definitely shifted since the {\it Chandra} HTEG 
observation in 2001. It is possible that a $k$T$_3$
$\sim$ 2.4\,keV component exists too. If so it may represent plasma
with nearly the same temperature found by S04. Further, the presence of 
a cool component with $k$T$_1$
$\sim$ 0.1\,keV is consistent with a value found by S04. This may or 
may not have the same cause as the wind-shocked fluxes emitted from 
other early B and Be stars. However, if so, it is a lower temperature than
expected. Only part (at most) of the $k$T$_2$ plasma could be produced
in the wind, and if so the large line widths evidenced in this component
indicate a larger turbulent broadening than is typical of winds in
other early B stars. 

\noindent {\it b)} The hot $k$T$_Q$ component appears stable and any
distribution of temperatures of individual sites around this mean value
must be small.

\noindent {\it c)} A subsolar abundance of iron is derived from the Fe\,XXV 
and Fe\,XXVI\,Ly$\alpha$ features, in agreement with several other 
determinations. As S04 also found, this Fe$_K$ abundance is 
significantly different of the abundance found from Fe-L ion lines. 
We have also discovered that [Fe$_K$] changes with time
and was significantly higher in 2004.

\noindent {\it d)} The light curve of this star again shows ubiquitous,
rapid flaring. This was 
also found by \citet{{Lopes07}} in \hd. However, unlike \hd, color
changes occurred only seldomly, and evidently not at all on the few hour
timescale noted in \hd.

\noindent {\it e)}  Our light curve shows a quasi-periodic lull every
2.8\,hours in 2004, similar to the cyclical lulls of 3.5 hours, 7-7.5 hours, 
and 5.8 hours noted by RSH02. These appear to occur in most epochs.

\noindent {\it f)} A thick absorption column affected 25\% of the hot
component in August 2001, but it had disappeared by February 2004.

\noindent {\it g)} Apparently variable Fe\,K and perhaps Si\,K 
fluorescent features are present. These emissions correlate with 
an absorption column (point {\it f}) that attenuates soft X-rays.

\noindent {\it h)} Broadening of warm ($k$T$_2)$ component 
lines was noticeable at both epochs but may have increased marginally from 
2001 ($\sim$ 0.4\,keV) to 2004 ($\sim$ 0.6\,keV).

\noindent {\it i)} The strengths of lines of two ions each of N and
Ne are underpredicted for XSPEC models with solar abundances.
Upon consideration of alternative explanations, we have interpreted 
these as evidence of abundance enhancements. However, we do not
understand the cause of these enhancements.

\noindent {\it j)} Given the identification of the Lyman\,$\beta$
feature in a few ions formed in the warm and cool plasmas,
there is more than a hint of nonequilibrium processes
in the environment of \gcas.~  This suggests formation in an (at most)
intermediate density environment, which is clearly separate from the
very high density plasma in which the flares are formed (Smith, Robinson,
\& Corbet 1998).

\noindent {\it k)} The unabsorbed flux of \gcas\ at 0.2--12 keV in 2004 from XMM-{\it Newton} ($\sim$ 1.7--2.8$\times$10$^{-10}$ erg\,s$^{-1}$) is consistent with the value observed in 2001 from Chandra ($\sim$ 2--3.1$\times$10$^{-10}$ erg\,s$^{-1}$).

In several of these respects the X-ray behavior of \gcas\ 
seems different from the one other analog studied in depth, \hd. 
In keeping with the differences of \hd, and indeed even the 
changing properties with time,  we refer to this star as 
having ``personality" - hence the title of this paper.

  Several changes we have noted are as remarkable as they 
were unexpected. Most especially, we have noted changes in the geometry 
of the circumstellar environment as reflected in the disappearance 
of one of the two columns, and affecting $\sim$ 25\% of the hard emission. 
The strong attenuation of the 2001 soft X-ray spectrum was clearly evident.
In addition, the strengths of the K fluorescence emission features
decreased in line with the decrease of the column absorption, suggesting that 
the part of the emission is caused by scattering of hard photons through the attenuating column that was present at the earlier time.

The reduced attenuation of the soft X-rays partially accounts for the
improved ability to study the nature of the warm and cool plasma emissions.
As a result, it is finally clear that these components are almost monothermal 
and therefore are not part of an integral structure with a smoothly varying 
thermal emission measure, such as an accretion column or a cooling flow plasma. 
Moreover, the broadening of the lines has increased during the 2001--2004 
interval, either because of an increase in a quasi-turbulence or the 
splitting of a former single region into two with different projected 
radial velocities.

In contrast to the warm component, a study of the variations in the
Fe\,K-shell lines with temperature shows that the hot plasma need not
consist of a uniform $k$T$_Q$ value. Rather, we believe that small 
variations in the hardness that are occasionally observed in our data 
and those discussed by SRC98, RS00, and RSH02 can be understood 
as variations of the instantaneous average temperature resulting from the
rapid evolution of small number of flares and nearby basal emission regions. 

The unique properties of each of the thermal components, the changing
absorption column geometry, and increases in line broadening all provide 
new hints to the mechanism responsible for the X-ray production. 
Although on one hand, the disappearance of the strong absorption column 
and the lack of correlation between the fluorescence features no
longer support the argument that the gas in the Be star's circumstellar
disk strongly interact with the hard X-rays, the discreteness of the
plasma components argues that 
they are likely to occupy distinct volumes. 

Future generations of X-ray telescopes such as the International 
X-ray Observatory (IXO) will be important in refining our
understanding of the spatial distribution of emission volumes of 
\gcas\ and its analogs. Their observations promise to address such 
question as whether the observed line broadening can be tied to the 
rotational velocity of the Be star and to resolving distinct sources, 
such as corotating active regions.

\begin{acknowledgements}

R.L.O. acknowledges financial support from the Brazilian agency FAPESP (Funda\c c\~ao de Amparo \`a Pesquisa do Estado de S\~ao Paulo) through a Postdoctoral Research Fellow grant (number 2007/04710-1).
We gratefully acknowledge the XMM-{\it Newton} User Support Group, in particular Nora Loiseau, Jan-Uwe Ness, and Matteo Guainazzi, for their help with problems on SASv8.0.1.

\end{acknowledgements}

\end{document}